\documentstyle[aps,prl,multicol,epsfig,psfrag]{revtex}
\begin{document}
\draft
\title{
A Neural Network Model for Apparent Deterministic Chaos in
Spontaneously Bursting Hippocampal Slices}

\author{B. Biswal${}^{1,2}$ and C. Dasgupta${}^{3,4}$} 

\address{
${}^1$ICA1, University of Stuttgart, Pfaffenwaldring 27, D-70569 Stuttgart,
Germany\\
${}^2$Department of Physics, Sri Venkateswara College, University of
Delhi, New Delhi, India \\
${}^3$Department of Physics, Indian Institute of Science,
Bangalore - 560 012, India\\
${}^4$Condensed Matter Theory Unit, JNCASR, Bangalore - 560 064, India}

\date{\today}

\maketitle
 
\begin{abstract}                 
A neural network model that exhibits stochastic population bursting is
studied by simulation. First return maps of inter-burst intervals
exhibit recurrent unstable periodic orbit (UPO)-like trajectories
similar to those found in experiments on hippocampal slices.
Applications of various control methods and surrogate analysis for
UPO-detection also yield results similar to those of experiments. Our
results question the interpretation of the experimental data as
evidence for deterministic chaos and suggest caution in the use of
UPO-based methods for detecting determinism in time-series data.

\end{abstract}
 
 

\begin{multicols}{2}
\narrowtext

With recent advances in the theory of dynamical systems~\cite{KG98} 
and nonlinear time-series analysis~\cite{KS97}, many recent studies
have looked for signatures of deterministic chaos in brain
dynamics~\cite{book}.  
This is expected to help in understanding brain functions in terms 
of the collective dynamics of neurons and in achieving short-term 
prediction and control with potential medical applications.

An experiment~\cite{Schiff94} on
spontaneously bursting rat hippocampal slices bathed in a
high-potassium medium has elicited much
interest~\cite{Glanz,CC95,So97,So98} in this
context. In this experiment, the occurrence of recurrent
trajectories resembling unstable periodic orbits (UPOs) in the first
return map of inter-burst intervals was interpreted as evidence for
chaotic dynamics. Application of chaos control~\cite{OGY90} was found to
be more effective than periodic pacing in making the bursts regular. An
anticontrol method designed to make the bursts more irregular was also
tested with positive results. The apparent success of these
control methods was interpreted as further evidence for the occurrence
of deterministic chaos. However, in subsequent work~\cite{CC95}, 
similar recurrent UPO-like trajectories were also
observed in simulations of a stochastic single-neuron model and
comparable success in chaos control and anticontrol was reported.
Recent studies~\cite{So97,So98} in which surrogate methods are used
to assess the statistical significance of UPOs claim to have 
re-established the original conclusion of deterministic chaos in 
hippocampal burst dynamics.

We have developed and studied by simulations a simple but biologically
plausible neural network model for hippocampal burst dynamics. While
the variability of the inter-burst intervals in our model arises from
stochastic transitions between two (quasi)attractors of the network
dynamics, our simulations and surrogate analysis of the data yield
results very similar to those reported~\cite{Schiff94,So97,So98} for
the brain slice experiment. Our work, thus, suggests that our model is
appropriate for describing the experimental system and casts doubt on
the interpretation of the experimental data as evidence for
deterministic chaos.

The neural network model we consider is based on one
developed\cite{MDU93} for the process of ``kindling'' of epilepsy.
The network consists of $N$ excitatory neurons represented by the
binary variables $\{S_i\}, i=1,2,.....,N$, with $S_i=0 (1)$ 
corresponding to quiescent (firing) states. The 
inhibitory neurons are collectively modeled by a background 
inhibition assumed to be proportional to the instantaneous number of
firing excitatory neurons. The ``local field'' $h_i(t)$ at 
the $i$th neuron at discrete time $t$ is given by

\begin{eqnarray}
h_i(t) & = & \sum_{j=1}^{N}[\{J_{ij}S_j(t)-wS_j(t)\} \nonumber\\
       & ~ & +\lambda\{K_{ij}S_j(t-\tau)-wS_j(t-\tau)\}],
\label{localfield}
\end{eqnarray}

\noindent where $w$ is the relative strength of inhibition, $\lambda$ 
is the relative strength of the delayed signal, and $\tau$ is the time
delay associated with the delayed signal. Starting from left, the four 
terms in Eq.(\ref{localfield}) respectively represent fast recurrent
excitation, fast global inhibition, slow recurrent excitation and slow 
global inhibition. These four ingredients are known\cite{Traub1} to be
essential for modeling hippocampal dynamics. Time $t$ is measured in
units of ``passes'', each pass 
corresponding to {\it random sequential updating} (RSU) of $N$ 
neurons according to the rule: $S_i(t+1)=1$ if $h_i(t) \ge 0$ and 
$S_i(t+1)=0$ if $h_i(t) < 0$. 

Initially a fixed number $(q)$ of random low-activity patterns 
(``memories'') $\{ \xi ^\mu _i \},~ i=1,2,.....,N;~ \mu=1,2,.....,q$, 
are stored in the synaptic matrices $J_{ij}$ and $K_{ij}$ through 
\begin{equation}
J_{i\ne j}= \Theta(\sum_{\mu=1}^{q}\xi_i^\mu \xi_j^\mu),\,\,
K_{i\ne j} = \Theta(\sum_{\mu=1}^{q}\xi_i^{\mu+1} \xi_j^\mu),
\label{kij}
\end{equation}
with $\xi_i^{q+1} = \xi_i^1$, $J_{ii}=K_{ii}=0$, and $\Theta(m)=1(0)$
for $m>0(m\le 0)$. The net activity of each memory, 
$ \sum_i \xi_i^\mu$, is set at a value $p<<N$. Our simulations are
carried out with the following parameter values: 
$N=200,~ q=20,~ p=10,~ w=0.6,~ \lambda=2,~ \tau=2$ passes.

\begin{figure}
\epsfig{bbllx=0,bblly=0,bburx=405,bbury=380,figure=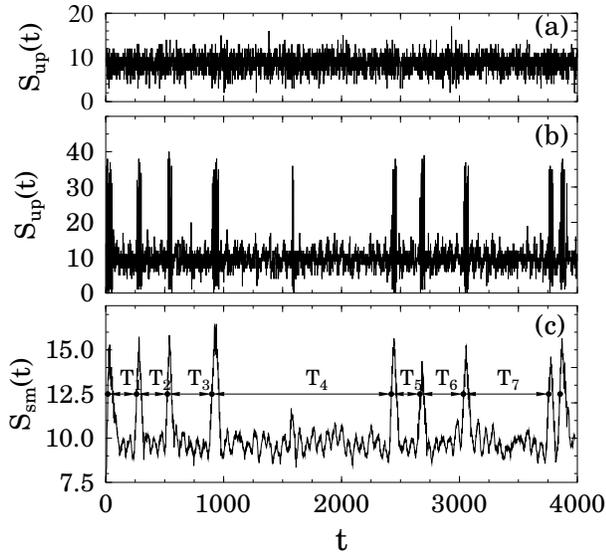,width=8cm}
\caption{(a) Network dynamics before kindling; (b) Kindled network dynamics
with spontaneous bursting; (c) Smoothened network activity with spikes. 
Thresholds are denoted by dark circles and $T_1, T_2,\ldots$ are the IBIs.}
\label{activity}
\end{figure}

The fast synapses $J_{ij}$ act to stabilize the system in a memory
state, the global inhibition prevents simultaneous activation of
multiple memories and the slow synapses $K_{ij}$ tend to move the
network from one memory to the next one after $\tau$ passes. For
$\lambda >1$, the network evolution follows a limit cycle in which all
the $q$ memories are visited sequentially with the net activity 
$S_{up}(t) = \sum_{i=1}^{N}S_i(t)$ showing
a low-amplitude oscillation around the average value $p$ 
[see Fig.\ref{activity}a]. To simulate the effects of the 
high-potassium medium, the network is
``chemically kindled'' through a Hebbian learning mechanism\cite{MDU93}
switched on during $50$ initial passes under reduced inhibition ($w$ is
reduced from $0.6$ to $0.24$). During this period, if $S_i=1$ and
$S_j=1$ more often than $6$ times over $10$ consecutive passes, then
the fast synaptic connection $J_{ij}$ is permanently set to 1.
Generation of these new excitatory synapses creates a
new high-activity (``epileptic'') attractor of the network dynamics.

The kindled network makes occasional, spontaneous, short-lived
excursions to the high activity attractor [see Fig.\ref{activity}b],
similar to the spontaneous population bursting observed in slice
experiments.  The net activity is smoothened (``low-pass filtered'') to
$S_{sm}(t) = \frac{1}{t_{sm}}\sum_{t'=t-t_{sm}+1}^{t}S_{up}(t')$, with
$t_{sm}=40$. The bursts appear as ``spikes'' in the smoothened data and
the inter-burst intervals (IBIs) are measured from positive threshold
crossings [see Fig.\ref{activity}c].  To prevent very rare
occurrences of persistent high activity, the network is reset to a
randomly chosen low-activity memory state whenever $S_{sm}(t)$ remains
continuously above the threshold for more than $20$ passes. 

The variability of the IBIs in our simulation arises entirely from the
{\it stochasticity of the sequence in which the neurons are updated}.
If a ``deterministic'' scheme in which the neurons are 
always updated in a (randomly chosen) {\it fixed} sequence is used, 
then the network remains in the low-activity state for most choices of
the sequence, and exhibits {\it periodic} transitions to the 
high activity attractor for certain rare sequences. Similar
behavior is found for parallel updating where all the neurons are
updated simultaneously. We note that the RSU scheme is the most 
appropriate one for modeling systems that do not have any
``clock-like'' mechanism.

\begin{figure}
\epsfig{bbllx=0,bblly=0,bburx=380,bbury=340,figure=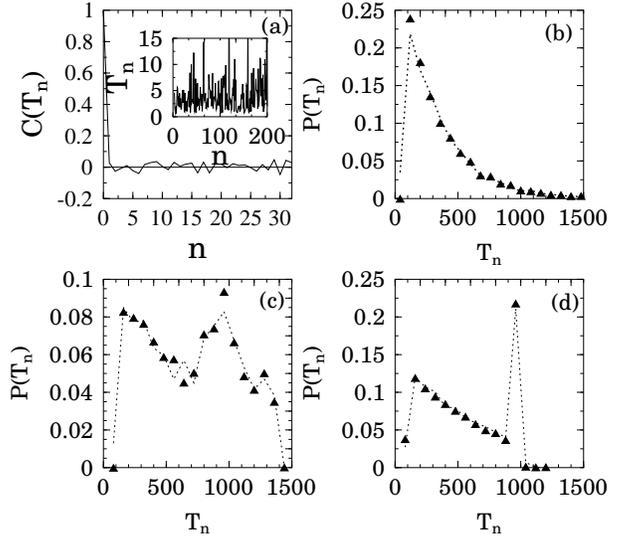,
width=8cm}
\caption{(a) Autocorrelation of IBIs and sample IBI data in units of 
$10^2$ passes (inset) for
network 1 in the absence of control. 
(b)-(d) Probability distribution of IBIs from network 
(dotted line) and Poisson (dark triangle) simulations for no control
(network 1, panel b), chaos control (network 6, panel c) and periodic
pacing (network 6, panel d).}  
\label{comp}
\end{figure}  

We have studied 10 networks with different realizations of the memory
states. As shown in the autocorrelation plot of Fig.\ref{comp}a, the
IBIs are uncorrelated in time in all the networks. Further, the
probability distribution of the IBIs follows a Poisson form, $P(T_n)
\propto \exp(-aT_n)$, for $T_n>2t_{sm}$ (this cutoff arises from our
smoothening procedure) in all the networks [see Fig.\ref{comp}b]. 
These observations strongly suggest that the spontaneous transition 
of the network from the low-activity state to the high-activity one is a 
stochastic process that has a small probability $p=1-\exp(-a)$
of occuring at each time step. As shown in 
Fig.\ref{comp}b, the results of a Monte Carlo simulation of a Poisson
process with probability $p$ of occurrence at each time step are
identical to those of network simulation. 
Results of our studies of the divergence of two
trajectories starting from neighboring points in one- and
two-dimensional delay embeddings of the IBI time-series 
are consistent with the IBIs being independent random variables. 

Values of $p$ for different
networks exhibit a wide variation in the range $8\times 10^{-4}$ to
$3 \times 10^{-3}$. Similar ``sample-to-sample'' variations and absence of
correlations in the IBI time-series were reported in
Ref.\cite{Schiff94b} for brain slices very similar to those of
Ref.\cite{Schiff94}. The variability in the raw IBI data of
Ref.\cite{Schiff94b} is qualitatively similar to that in our simulation
data [see inset of Fig.\ref{comp}a]. 

\begin{figure}
\epsfig{bbllx=-100,bblly=0,bburx=330,bbury=280,figure=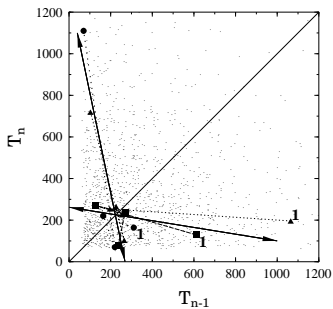,
width=6cm}
\caption{Recurrent UPO-like trajectories in network $1$. 
Three sequences (filled circles, squares and triangles, starting points
denoted by $1$) starting at
different times around the same stable and unstable manifolds (dark
lines) are shown.} 
\label{returnmap}
\end{figure}

Although the bursting in our model is stochastic, first return
maps of the IBIs show purely accidental
occurrences\cite{CC95} of UPO-like trajectories\cite{Schiff94} 
that approach a fixed point ($T_n=T_{n-1}=T^*$) along a line,
$T_n-T^*=s_1(T_{n-1}-T^*)$ (the stable manifold) and then diverge 
away along another line, $T_n-T^*=s_2(T_{n-1}-T^*)$ (the unstable manifold). 
In each of the 10 networks studied, many recurrent UPO-like trajectories 
that satisfy all the criteria adopted in \cite{Schiff94} are 
found. Typical trajectories used to identify an apparent UPO and the
associated manifolds are shown in Fig.\ref{returnmap}.

\begin{figure}
\epsfig{bbllx=0,bblly=0,bburx=400,bbury=150,figure=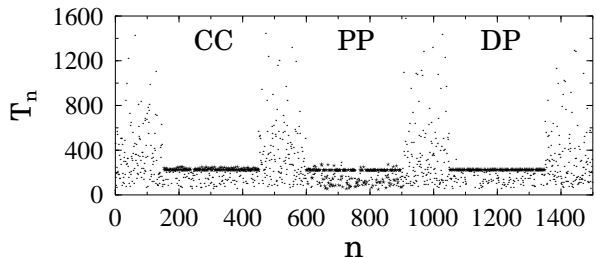,
width=8cm}
\caption{Results of chaos control(CC), periodic
pacing(PP), and demand pacing (DP) runs of $300$ IBIs each in network
$1$. No control is applied during the intermediate periods of $150$
IBIs. IBIs generated by external stimulation are denoted by stars.}
\label{control}
\end{figure}  

Following Ref.\cite{Schiff94}, we have studied the effects of various
control methods. In the {\it Proportional Perturbative Feedback} method
of chaos control (CC) \cite{Garfinkel92}, the system point is
repeatedly placed on the stable manifold of an UPO detected during a
trial period withour control. This is done by generating bursts at
appropriate times through external stimulation, modeled in our
simulation by a reduction of the inhibition strength $w$ from $0.6$ to
$0.24$ for $5$ passes. Given a burst with IBI $T_{n-1}$, the external
stimulation is applied $T_n = s_1(T_{n-1}-T^*)+T^*$ passes after the
burst if no spontaneous burst occurs before its application. In
periodic pacing (PP), the external stimulation is applied at fixed
intervals of $T^*$ irrespective of the occurrence of spontaneous
bursts, and in demand pacing (DP), the external stimulation is applied
$T^*$ passes after a burst only if no spontaneous burst occurs prior to
its application. 

Typical results of the application of control,
shown in Fig.\ref{control}, are qualitatively similar to those reported
in Ref.\cite{Schiff94}. All three methods make the
bursts more periodic by preventing the occurrence of large IBIs.  While
CC prevents the occurrence of natural IBIs above the stable manifold,
PP and DP eliminate IBIs larger than $T^*$. This is why CC is
effective when $T^*$ is small and the slope $s_1$ of the stable manifold
is close to zero. In such cases, CC works better than PP because the
occurrence of spontaneous bursts between two successive stimulations in
PP increases the number of short IBIs. Since this does not happen in
DP, this method yields the best results. For example, in network 1
for which the selected UPO,  shown in Fig.\ref{returnmap}, has 
$T^*=225$, $s_1=-0.162$, $s_2=-4.8 $, the fraction of
IBIs in the range $T^* \pm t_{sm}$, measured in 5000 IBI runs, is 0.17,
0.68, 0.52 and 0.69 for no control, CC, PP and DP respectively (the
statistical error bars for these fractions are less than 0.01). Thus,
CC and DP work better than PP (see Fig.\ref{control}) in this case.
The corresponding values for network 6, in which $T^*=937$, $s_1=-0.476$,
$s_2 =-1.68 $, are 0.03, 0.09, 0.21 and 0.4, indicating better performance
of PP and DP over CC (see Figs.\ref{comp}c, \ref{comp}d). 


\begin{figure}
\psfrag{T}{$\hat{T}$}
\epsfig{bbllx=15,bblly=15,bburx=350,bbury=180,figure=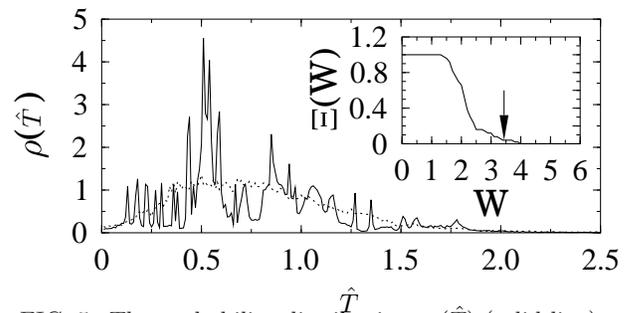,width=8cm}
\caption{The probability distributions $\rho(\hat{T})$ (solid line)
and $\bar{\rho}_{sur}(\hat{T})$ (dots) for a 128-IBI dataset. Inset: the
function $\Xi(W)$ used for estimating the statistical significance of
UPOs. The arrow marks the maximum value of 
$[\rho(\hat{T})-\bar{\rho}_{sur}(\hat{T})]$.}
\label{ps}
\end{figure}  

Even when CC works 
well, 70-90\% of the IBIs near $T^*$ are actually generated by the
stimulation (see Fig.\ref{control}): the probability of occurrence of
spontaneous IBIs near $T^*$ is not significantly increased during
control. A similar behavior in the experiment is suggested by the
experimental observation\cite{Schiff94} that intermittent applications
of the external stimulation {\it does not} produce effective control.
Increasing the strength of stimulation improves control in our
simulations by preventing rare failures of the stimulation to generate
the required burst. This is qualitatively similar to the results of
double-pulse control in the experiment. The anticontrol method proposed
in Ref.\cite{Schiff94} fails if applied over long durations ($ \sim
1000$ IBIs). In applications of smaller duration ($\sim 100$ IBIs), the
variability of the IBIs increases in some cases due to statistical
fluctuations. In the experiment, anticontrol was applied for short
durations and failed in most cases.

Our control results are similar to those reported in Ref.\cite{CC95}
for a stochastic single-neuron model. We have also simulated the
effects of control on simple Poisson processes with $p$ determined from
the network data. The close agreement (see Figs.\ref{comp}c,
\ref{comp}d) between the network simulation results
and those of our Poisson simulations firmly establishes the stochasticity
of the bursting in our model.
 
Recently, the statistical significance of UPOs in brain slice data has
been examined\cite{So97,So98} using surrogate analysis 
of windowed IBI datasets. These studies use a {\em dynamical 
transformation}\cite{So97} that maps the neighborhood of an UPO
in the return map to the UPO itself. First, the IBI time-series $\{T_n\}$ is 
transformed by
\begin{equation}
\hat{T_n}~=~\frac{T_{n+1}-s_n(k)T_n}{1-s_n(k)}
\label{pstransform}
\end{equation}
\noindent where 
$s_n(k)~=~[(T_{n+2}-T_{n+1})/(T_{n+1}-T_n)]+k(T_{n+1}-T_n)$.
The probability distribution $\rho(\hat{T})$ of the transformed time-series 
$\{\hat{T_n}\}$ is expected to show a sharp peak at an UPO, $\hat{T}=T^*$.
Spurious peaks are eliminated\cite{So97} by averaging over many random
values of $k$. Using many realizations of 
Gaussian-scaled phase-shuffled surrogates\cite{Theiler92} of the
original time-series, the probability that the observed peaks in
$\rho(\hat{T})$ could be modeled by the surrogates 
is estimated in the following way. From the collection of
\{$\rho_{sur}(\hat{T})$\}, the ensemble average $\bar{\rho}_{sur}(\hat{T})$
is obtained. The deviation of $\rho_{sur}(\hat{T})$ for a particular 
surrogate from the mean $\bar{\rho}_{sur}(\hat{T})$  is measured by
\begin{equation}
w(\hat{T})=\rho_{sur}(\hat{T})-\bar{\rho}_{sur}(\hat{T}).
\end{equation}
For each of the surrogates, $W=max(w(\hat{T}))$ is measured
and the fraction $\Xi(W')$ of surrogates with maximum deviation
$W>W'$ is found. The statistical significance of a
candidate UPO at $\hat{T}^*$ is estimated from the value of $\Xi(W)$ at
$W=\rho(\hat{T}^*) - \bar{\rho}_{sur}(\hat{T}^*)$. In our analysis, we
used 100 surrogates, $k=5R$ where $R$ is a random number in $[1,-1]$,
and 500 values of $R$. We find that this method
successfully rejects the UPO-like trajectories found in the network IBIs 
(properly rescaled to match the analysis of Ref.\cite{So97})
only for long datasets ($>1024$ IBIs), but not for shorter datasets
(windows). Results for a typical $128$-IBI window with a 
``statistically significant'' UPO are shown in Fig.\ref{ps}. 
In this case, the probability of finding the peak of $\rho(\hat{T})$
(near $\hat{T}=0.5$, with $W=3.437$) 
in the surrogates is close to $4\%$, leading to the (false)
identification of a statistically significant UPO at $95\%$ confidence
level. In such analysis of many non-overlapping windows of small width 
from $10$ networks (544, 272 and 136 windows of 32, 64 and 128 IBIs
respectively), statistically significant (at the $95\%$ confidence
level) UPOs are found in $9-10\%$ of the windows. The number of
networks for which such statistically
significant UPOs are present in at least one of the windows is 8, 10 
and 5 for window size $32$, $64$ and $128$, respectively. These numbers
are very similar to those reported in Ref.\cite{So98}. Very similar
results were found (e.g. 75 out of 1024 non-overlapping 32-IBI windows
showing statisically significant UPOs for network 1) in a surrogate
analysis using a two-dimensional delay embedding~\cite{So97} and both
overlapping and non-overlapping windows. Ref.\cite{So98} reports the
detection of statistically significant UPOs of period higher than one
in {\it intracellular} data which reflect a combination of collective 
and single-neuron dynamics. Our network with binary neurons 
is not appropriate for modeling such data. Nevertheless, a similar 
analysis of
our IBI data is planned in future work.

In conclusion, our neural network model 
with stochastic bursting dynamics
successfully reproduces most of
the findings of the brain slice experiment. This result
casts doubt on the interpretation of the experimental data 
as evidence for deterministic chaos.
Our work also shows that determinism and
stochasticity need not be mutually exclusive: the evolution of our
network in the low- and high-activity states is mostly deterministic,
while the transitions between these two states are stochastic. 

We thank Dr. G. R. Ullal for useful discussions.

\end{multicols}  
\end{document}